\documentclass[12pt]{iopart}

\usepackage{graphicx}
\usepackage{braket}
\usepackage{color}
\usepackage{iopams}

\newcommand{\unit}[1]{\ensuremath{\, \mathrm{#1}}}

\DeclareMathAlphabet\mathbfcal{OMS}{cmsy}{b}{n}

\begin{document}

\title{Non-perturbative laser effects on the electrical properties of
graphene nanoribbons}

\author{Hern\'{a}n L Calvo$^1$, Pablo M P\'{e}rez Piskunow$^2$, Horacio
M Pastawski$^2$,  Stephan Roche$^{3,4}$, and Luis E F Foa Torres$^2$}

\address{$^1$ Institut f\"{u}r Theorie der Statistischen Physik, RWTH
Aachen University, 52056 Aachen, Germany and JARA - Fundamentals of
Future Information Technology}

\address{$^2$ Instituto de F\'{\i}sica Enrique Gaviola (IFEG-CONICET)
and FaMAF, Universidad Nacional de C\'{o}rdoba, Ciudad Universitaria,
5000 C\'{o}rdoba, Argentina}

\address{$^3$ CIN2 (ICN-CSIC), Catalan Institute of Nanotechnology,
Universidad Aut\'{o}noma de Barcelona, Campus UAB, 08193 Bellaterra
(Barcelona), Spain} 

\address{$^4$ Instituci\'{o} Catalana de Recerca i Estudis Avan\c{c}ats
(ICREA), 08070 Barcelona, Spain}

\ead{lfoa@famaf.unc.edu.ar}
 
\begin{abstract}
The use of Floquet theory combined with a realistic description of the
 electronic structure of illuminated graphene and graphene nanoribbons
 is developed to assess the emergence of non-adiabatic and
 non-perturbative effects on the electronic properties. Here, we
 introduce an efficient computational scheme  and illustrate
 its use by applying it to graphene nanoribbons in the presence of both
 linear and circular polarization. The interplay between confinement due
 to the finite sample size and laser-induced transitions is shown to
 lead to sharp features on the average conductance and density of
 states. Particular emphasis is given to the emergence of the bulk limit
 response.
\end{abstract}

\pacs{73.23.-b, 72.10.-d, 73.63.-b}

\maketitle

\section{Introduction}

Carbon-based materials such as graphene and carbon nanotubes constitute
a privileged family of novel nanomaterials with outstanding properties
\cite{Geim2009, CastroNeto2009, Dubois2009}. Within this playground,
graphene optics and optoelectronics has become one of the most exciting
areas for research \cite{Bonaccorso2010, Xia2009, Karch2011}. Besides
being one of the best tools for non-invasive characterization and
probing of carbon-based materials \cite{Raman}, light can also be used
as means for achieving new functions such as improved energy harvesting
\cite{Gabor2011}, or novel plasmonic properties \cite{Koppens2011,
Chen2012} and there are big expectations that we are going to have even
more in the coming years. At the core of these phenomena we are
confronted with fundamental aspects of the interaction between light and
matter in low dimensionality. 
    
While the effects of light and other bosonic excitations on the
electronic properties is usually treated within a Fermi Golden rule
and/or adiabatic approximations, their breakdown is indeed a possibility
in the graphene flatland \cite{FoaTorres2006, Lazzeri2006}, as
demonstrated by some remarkable experiments \cite{Pisana2007}.

The illumination with a laser was also proposed to lead to
non-perturbative and non-adiabatic effects in graphene including the
opening of a bandgap \cite{Syzranov2008} owing to photon-assisted
coupling between electronic states at half the photon energy,
i.e. $\pm \hbar \Omega/2$. Further studies pointed out that circularly
polarized light may also lead to a Hall effect without a static magnetic
field: the photovoltaic Hall effect \cite{Oka2009} which lacks an
experimental confirmation.

These puzzling possibilities attracted much attention \cite{Calvo2011,
Zhou2011, Savelev2011} and recent
atomistic simulations of the dc electrical response hint that a laser in
the mid-infrared would be optimal for experimental verification
\cite{Calvo2011}. Further studies focused on the optical response
\cite{Zhou2011, Busl2012} as well as other interesting issues
\cite{Kibis2010, Iurov2012, Liu2012, San-Jose2012}. Recently, the
possibility of inducing topologically protected states by laser
illumination in semiconductors \cite{Lindner2011} and both in monolayer
\cite{Kitagawa2011, Gu2011} and bilayer \cite{SuarezMorell2012} graphene
has become a hot field  \cite{Cayssol2012,Rudner2012} with an impact on
other areas such as condensates \cite{Crespi2013} and photonic crystals,
where experiments are already available \cite{Rechtsman2012}. 

Most of the results mentioned before have been derived for \textit{bulk}
monolayer graphene (bilayer graphene was also studied in
\cite{Abergel2009, SuarezMorell2012}) and only few results are available
for the case of \textit{graphene nanoribbons} \cite{Gu2011,
Calvo2012}. In this paper, we extend over our previous results
\cite{Calvo2012} giving a more comprehensive and in-depth view of the
numerical scheme used and the interplay between lateral confinement and
photon-assisted processes in illuminated graphene nanoribbons. To this
end, we apply Floquet theory to a nearest neighbor tight-binding model
to describe the $\pi$-orbitals around the Fermi energy in presence of a
time-periodic field. In this approach, the average density of states
(DOS) and the dc component of the conductance are calculated through an
efficient reduction of the Floquet Hamiltonian by a recursive decimation
procedure \cite{Pastawski2001}. The power of this technique allows us to
explore the electrical properties of the system in a wide range of
parameters, including arbitrary edge geometries and ribbon sizes as well
as different laser polarizations.

As will be made clear in the next sections, the non-perturbative
character of the field is revealed through a strong dependence in the
laser-induced gaps with respect to the direction of the polarization. In
addition to the known dynamical gaps developed at $\pm \hbar \Omega/2$
\cite{Syzranov2008}, further modifications in the DOS emerge as a
consequence of the inter-mode coupling induced by the laser. The effect
is discussed in both armchair and zigzag nanoribbons for different
orientations of the radiation field. Finally, for large nanoribbons the
effects of both linear and circular polarizations are also discussed. We
show how the above modifications in the DOS average out around $\pm
\hbar \Omega /2$, thereby recovering the known solution in the bulk
limit \cite{Syzranov2008, Oka2009}.

Our work is organized as follows. In section \ref{sec:model} we
introduce the considered model and give a detailed overview of the
simulation scheme. Section \ref{sec:results} is devoted to the
discussion of our results. Finally, we present our conclusions in
section \ref{sec:conclusions}.

\section{Simulation scheme: Floquet theory applied to illuminated
 graphene nanoribbons} 
\label{sec:model}

In this section we outline the approach used to investigate the effects of
the laser field in the electronic structure and transport properties of
graphene nanoribbons. Our starting point is the description of the
$\pi$-orbital electrons through the following tight-binding Hamiltonian
\begin{equation}
\hat{\mathcal{H}}(t) = \sum_i \epsilon_i \hat{c}_i^\dagger \hat{c}_i -
 \sum_{\braket{i,j}} \gamma_{ij}(t) \hat{c}_i^\dagger
 \hat{c}_j, 
\end{equation}
where $\hat{c}_i^\dagger (\hat{c}_i)$ creates (annihilates) an electron
at site $i$ and the sum in the second term only takes pairs of
nearest-neighbor sites. The potential energy induced by, e.g., a gate
voltage or local impurities, is represented by the on-site energies
$\epsilon_i$ while the hopping term $\gamma_{ij}$ gives the transition
amplitude between sites $i$ and $j$.

The time-dependent field is considered by neglecting the small magnetic
component and with an electric field which is written in a Weyl's gauge
in terms of a vector potential: $\mathbf{E}=-\partial\mathbf{A}/\partial
t$. The vector potential is included through the Peierls substitution
\cite{Peierls1933}, which introduces an additional phase in the hopping
amplitude connecting two adjacent sites $i$ and $j$, i.e.
\begin{equation}
\gamma_{ij}(t) = \gamma_0 \exp \left[ i \frac{2\pi}{\Phi_0}
\int_{\mathbf{r}_j}^{\mathbf{r}_i} \mathbf{A}(t) \cdot d\mathbf{r}
\right],
\label{eq:peierls}
\end{equation}
where $\gamma_0 = 2.7\unit{eV}$ is the typical hopping amplitude at zero
field \cite{Dubois2009} and $\Phi_0$ is the magnetic flux
quantum. For monochromatic waves, the gauge potential related to the
time-dependent electric field is defined as
\begin{equation}
\mathbf{A}(t) = A_0 \left[\cos \beta \cos (\Omega t)
\mathbf{x} + \sin \beta \cos (\Omega t-\varphi) \mathbf{y}\right],
\label{eq:vecpot}
\end{equation}
where $A_0 = E_0/\Omega$ and $E_0$ is the amplitude of the	
electric field. The direction and polarization of the field are fixed by
the parameters $\beta$ and $\varphi$, respectively. In the next section
we will concentrate on three paradigmatic cases, namely, linear
polarization along either the (longitudinal) $x$-axis or (transversal)
$y$-axis and circular polarization.

To deal with the time-dependence in the electronic hopping terms, we use Floquet theorem \cite{Shirley1965, Kohler2005}. In the next
subsection, the Floquet Hamiltonian accounting for the
$2\pi/\Omega$-periodic real-space Hamiltonian is discussed in detail
together with the employed technique for the derivation of the
observables (DOS and conductance) which are investigated.

\begin{figure}[tbp]
\includegraphics[width=8.5cm]{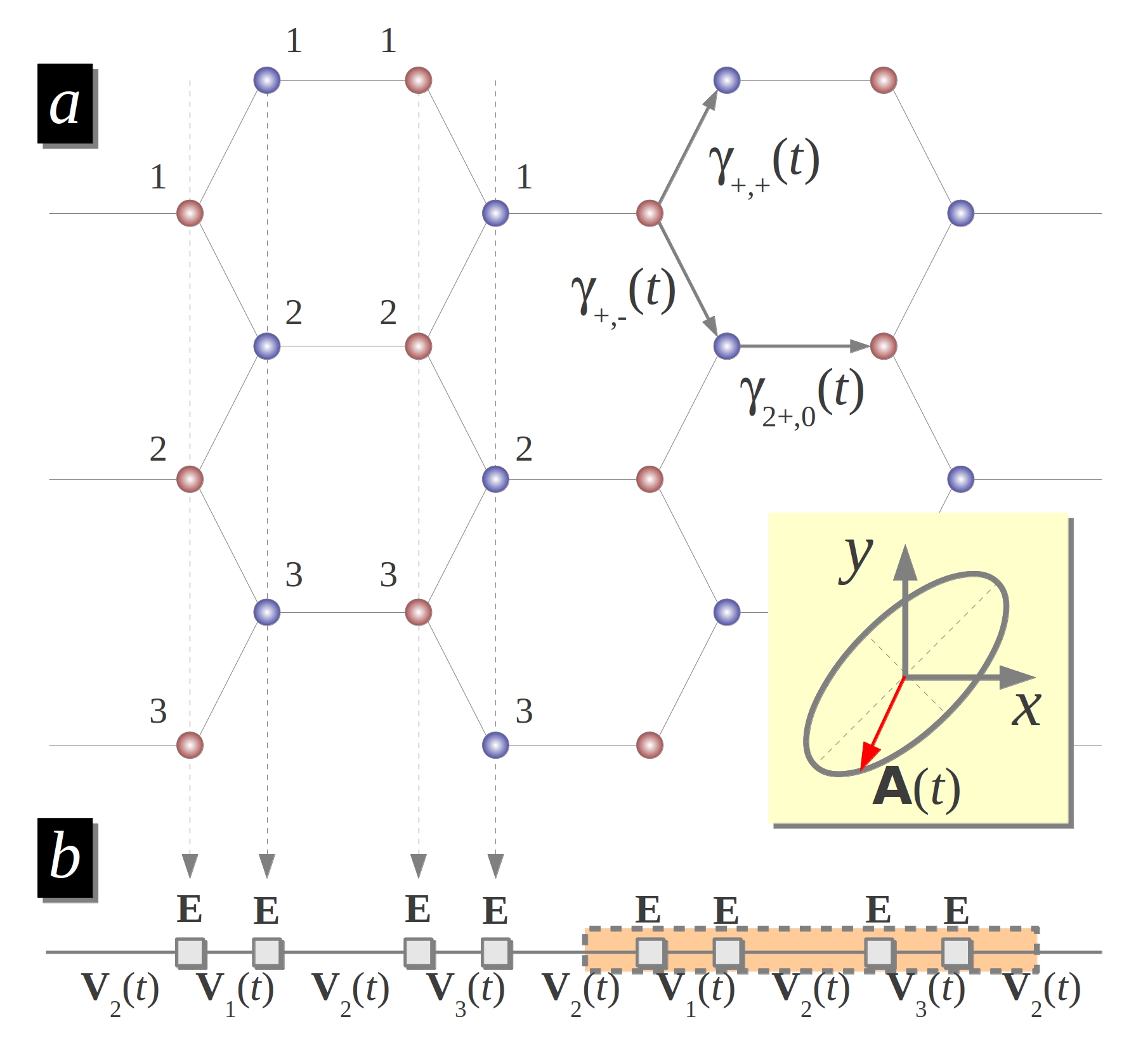}
\caption{(a) Scheme of the considered model for the example of an
 armchair nanoribbon. The laser field is applied perpendicular to the
 lattice and defines the time-dependent hoppings $\gamma_{+,+}(t),
 \gamma_{+,-}(t)$ and $\gamma_{2+,0}(t)$. (b) Representation of the
 real-space Hamiltonian in terms of diagonal blocks $\mathbf{E}$ and
 hopping matrices $\mathbf{V}_i(t)$. The unit cell for the case of an
 armchair nanoribbon is indicated by a rectangle.}
\label{fig:scheme}
\end{figure}

\subsection{Determination of the matrix elements of the Floquet Hamiltonian}

A scheme of the considered system is shown in figure \ref{fig:scheme}(a), where
the phases of the hopping amplitudes account for the time-dependent
vector potential associated to the laser. For two adjacent (transversal)
arrays in the lattice, we consider the hopping amplitudes always going
from left to right. With the assumption that the electric field is
uniform along the whole sample, the phases in equation \ref{eq:peierls} are
given by the scalar product between the vector potential and the vector
$\mathbf{r}_{ij} = \mathbf{r}_i - \mathbf{r}_j$ connecting the two
sites. This allows to fully describe the system in terms of three
possible orientations of the vector $\mathbf{r}_{ij}$,
i.e., $\mathbf{r}_{ij} = a(\cos \alpha_{ij}, \sin \alpha_{ij})$, with $a
\simeq 0.142\unit{nm}$ the distance between nearest-neighbor carbon
atoms and $\alpha_{ij} = 0,\pm \pi/3$. For these orientations, the
hopping elements are
\begin{eqnarray}
\gamma_{2+,0}(t) &=& \gamma_0 e^{i 2 z_x \cos(\Omega t)},\\
\gamma_{+,\pm}(t) &=& \gamma_0 e^{ i (z_x \pm z_y \cos
 \varphi)\cos(\Omega t)} e^{\pm i z_y \sin \varphi \sin(\Omega t)},
\end{eqnarray}
respectively, where the indices are now related to the $x$ and $y$
components of the vectors connecting the carbon atoms. To simplify the
notation, we define adimensional variables $z_x = \pi a A_0 \cos
\beta/\Phi_0$ and $z_y = \sqrt{3} \pi a A_0 \sin \beta/\Phi_0$ which
quantify the relative strength of the laser.

We divide the real-space Hamiltonian in diagonal blocks $\mathbf{E}$
accounting for the on-site energies in the lattice which belong to the
same transversal array. In homogeneous samples, they are simply zero,
i.e. $\mathbf{E} = \mathbf{0}$. The hopping matrices $\mathbf{V}_i(t)$
connecting the arrays alternate periodically depending on the particular
position in the unit cell (see figure \ref{fig:scheme}(b)). The dimension
of these block matrices is given by the number $N$ of carbon atoms along
the transversal direction.

A Fourier decomposition of the time-dependent Hamiltonian spans the real
space into a composite space $\mathcal{R} \times \mathcal{T}$ including
the $2\pi/\Omega$-periodic functions \cite{Shirley1965,Kohler2005}. The
basis of this Floquet space is conformed by the states $\ket{i,n}$,
where the first index labels the site location of the sites and the
second, the Fourier index, indicates the number of photons. The
resulting time-independent (Floquet) Hamiltonian $\mathbf{H}_F$ is
determined by the hopping elements connecting the states $\ket{i,n}$ and
$\ket{j,n+m}$
\begin{eqnarray}
\gamma_{+,\pm}^m &=& \gamma_0 \sum_{k=-\infty}^{\infty} i^k J_k(z_x \pm
 z_y \cos \varphi) J_{m-k}(\pm z_y \sin \varphi),\\ 
\gamma_{2+,0}^m &=& \gamma_0 i^m J_m(2 z_x),
\end{eqnarray}
where $J_n(z)$ is the Bessel function of order $n$. In contrast to similar
treatments of the periodic time-dependent field based on the {\bf
\em{k.p}} approach \cite{Syzranov2008, Oka2009}, the present calculations may involve
transitions with more than a single photon. However, for the
mid-infrared regime ($ \hbar\Omega \simeq 140\unit{meV}$) and laser
power ($P = 1 - 10 \unit{mW/\mu m^2}$) considered here we have $z_x, z_y
\ll 1$, such that these transitions decay rapidly and the leading contributions still
come from the renormalization of the hoppings ($m = 0$) and inelastic
transitions with a single photon ($m = \pm 1$). The method is here
illustrated for the specific case of armchair nanoribbons, but the same
strategy can be easily adapted to other edge geometries.

Now that we have an explicit expression for the hopping amplitudes, we
can compute the Floquet Hamiltonian by noticing that each block in
figure \ref{fig:scheme}(b) is splitted into $2N_F+1$ ``replicas''
accounting for states with different Fourier index. In this sense, $N_F$
denotes the maximum number of considered Fourier modes and dictates
the truncation of the total Floquet space. The on-site energies in the
diagonal blocks are given as a multiple integer of the photon energy,
i.e., $\epsilon_{i,n} = n\hbar \Omega$, with $n =
-N_F,\dots,N_F$. According to this Fourier expansion, the structure of
the Floquet Hamiltonian can be understood as a block tridiagonal
matrix where each block is of dimension $(2N_F+1)N$. The diagonal blocks
$\mathbf{E}_F$ include the arrays with different Fourier index. This
group of arrays then form a layer and the off-diagonal blocks contain
the hopping amplitudes connecting them. Since in our representation
each layer contains carbon atoms belonging to the
same sublattice ($A$ or $B$), the matrix elements of the diagonal blocks
simply read $\bra{i,n}\mathbf{E}_F\ket{j,m} =  n \hbar \Omega
\delta_{ij}\delta_{nm}$. For off-diagonal blocks, however, we need to
distinguish three different hopping matrices $\mathbf{V}_{F,k}$, with
$k=1,2$ or $3$, according to which layers they interconnect. To account for
transitions with different number of photons, each matrix is divided, in
turn, into $2N_F+1$ submatrices such that
$\bra{i,n}\mathbf{V}_{F,k}\ket{j,n+m} = \bra{i}\mathbf{V}_{k}^m\ket{j}$,
where
\begin{eqnarray}
\mathbf{V}_1^m &=& \left(
\begin{array}{ccc}
\gamma_{+,+}^m & \gamma_{+,-}^m & \\
0 & \gamma_{+,+}^m  & \\
  &   & \ddots
\end{array} \right),
\mathbf{V}_2^m = \left(
\begin{array}{ccc}
\gamma_{2+,0}^m & 0 & \\
0 & \gamma_{2+,0}^m  & \\
  &   & \ddots
\end{array} \right), \\
\mathbf{V}_3^m &=& \left(
\begin{array}{ccc}
\gamma_{+,-}^m & 0 & \\
\gamma_{+,+}^m & \gamma_{+,-}^m & \\
  &  & \ddots
\end{array} \right).
\end{eqnarray}
This representation of the Floquet Hamiltonian constitutes our starting
point in the analysis of the electronic properties of the ribbons in the
presence of a time-dependent field. In previous works \cite{Calvo2011},
the particular case of armchair nanoribbons illuminated by linearly
polarized light along the longitudinal direction was intensively
studied. This was motivated by the fact that this geometry allows for a
convenient decomposition of the Floquet Hamiltonian into normal modes,
thereby making it a suitable model to analyze the bulk limit for a large
lateral size of the ribbon. In the next section, however, we will 
concentrate first on relatively small sized ribbons, and our efforts will
be devoted to investigate the role of the laser field in the
characteristics of the electronic structure due to lateral
confinement. After discussing the interplay between these two effects,
we will explore the bulk limit for different directions of the linear
polarization and also for circularly polarized fields.

\subsection{Density of states and Conductance}

We now comment on the quantities of interest that we want to address in
the next section when taking specific values for the size and edge
geometry of the ribbons as well as tuning the laser
field. Starting from the above mentioned Floquet Hamiltonian, it is
possible to define the retarded Floquet-Green functions according to
\cite{Martinez2003,FoaTorres2005} 
\begin{equation}
\mathbf{G}_\mathrm{F}(\varepsilon) = \left(\varepsilon \mathbf{1} -
\mathbf{H}_\mathrm{F}\right)^{-1}.
\end{equation}
\begin{figure}[tbp]
\includegraphics[width=8.5cm]{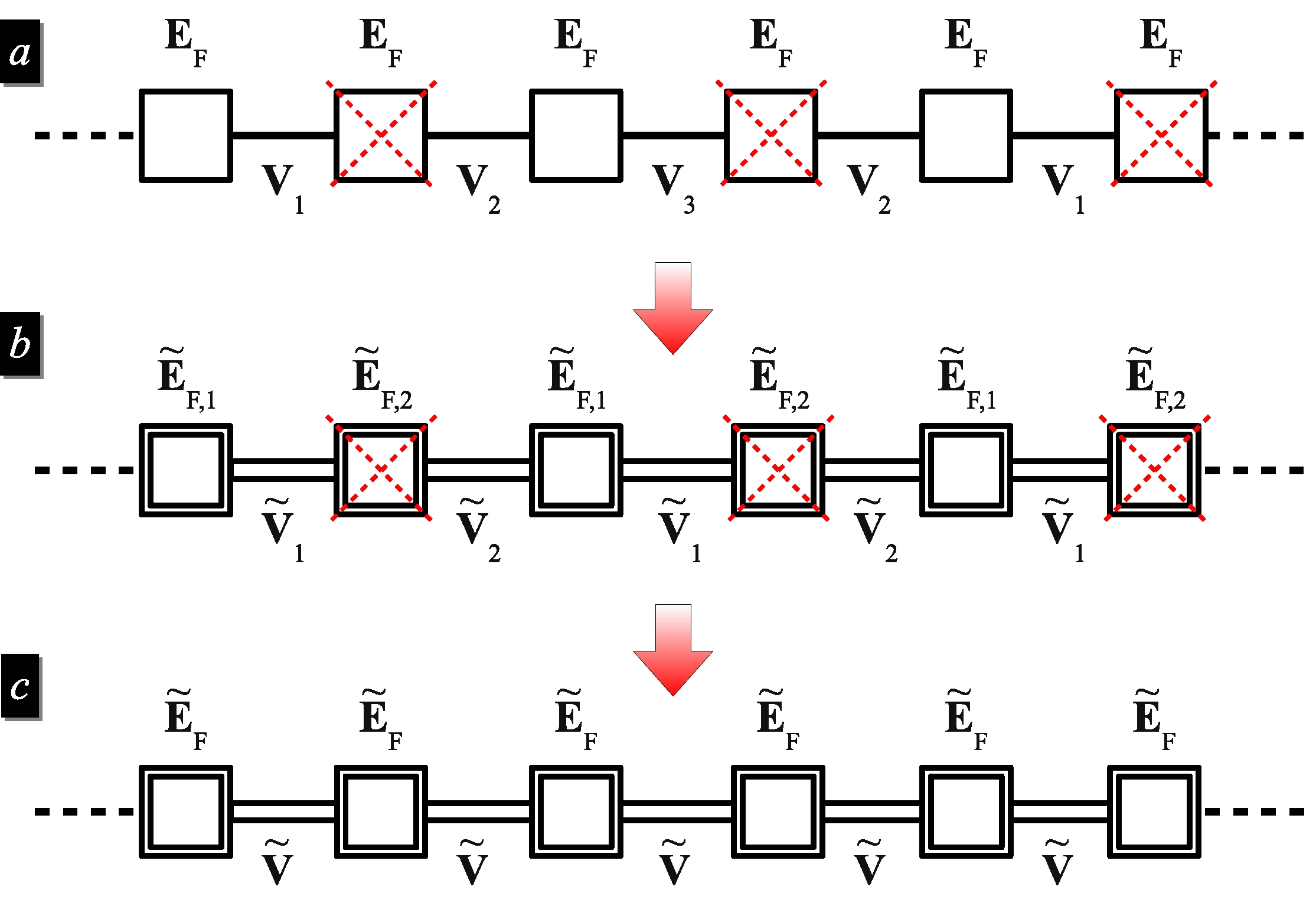}
\caption{Decimation procedure in Floquet space. (a) Effective
 representation of $\mathbf{H}_F$. Dashed red crossing lines denote the
 reduced blocks. (b) Effective Hamiltonian after one step: The diagonal
 block matrices $\tilde{\mathbf{E}}_{F,i}$ and the effective hopping
 matrices $\tilde{\mathbf{V}}_i$ are renormalized by the reduced
 blocks. (c) Homogeneous effective Hamiltonian after the second step of
 the decimation procedure.}
\label{fig:decimation}
\end{figure}
Since we want to analyze how the electronic properties of the ribbon are
affected by the laser, we refer to the average density of states (DOS)
obtained after tracing over the sites with zero photon, i.e.
\begin{equation}
\mathcal{N}_0(\varepsilon) = -\frac{1}{\pi} \lim_{\eta \rightarrow 0}
\mathrm{Im} \left[ \sum_{i=1}^N \bra{i,0} \mathbf{G}_F(\varepsilon+i\eta)
\ket{i,0} \right],
\label{eq:dos1}
\end{equation}
where the (imaginary) regularization energy $i\eta$ is carefully chosen
to reach the thermodynamical limit along the longitudinal direction. A
``brute-force'' calculation of the inverse of the Floquet Hamiltonian
could in principle represent a hard (or even impossible) task when
considering the full system. However, the decimation technique
\cite{Economou2006,Pastawski2001}, based on the recursive calculation of
the self-energy correction $\mathbf{\Sigma}_F(\varepsilon)$ to the
diagonal block matrices, constitutes an appropriate strategy to
circumvent this hurdle. Figure \ref{fig:decimation} shows a scheme where
the Floquet Hamiltonian is represented by an effective linear
chain. Here, the squares correspond to the diagonal blocks
$\mathbf{E}_F$, which are connected through the hopping matrices
$\mathbf{V}_i$. In this procedure, we calculate the effective
Hamiltonian resulting from the systematic elimination of blocks in the
lattice (dashed red crossing lines in the figure). This reduction of the
basis along the longitudinal direction leads to a renormalization of
both the diagonal blocks and hopping matrices. According to figure
\ref{fig:decimation}, after the first decimation step (panel $b$), these
read
\begin{eqnarray}
\tilde{\mathbf{E}}_{F,1} &=& \mathbf{E}_F + \mathbf{V}_1 \frac{1}{\varepsilon
 \mathbf{1} - \mathbf{E}_F} \mathbf{V}_1^\dag +\mathbf{V}_2^\dag
 \frac{1}{\varepsilon \mathbf{1} - \mathbf{E}_F} \mathbf{V}_2,\\
\tilde{\mathbf{E}}_{F,2} &=& \mathbf{E}_F + \mathbf{V}_3 \frac{1}{\varepsilon
 \mathbf{1} - \mathbf{E}_F} \mathbf{V}_3^\dag +\mathbf{V}_2^\dag
 \frac{1}{\varepsilon \mathbf{1} - \mathbf{E}_F} \mathbf{V}_2,\\
\tilde{\mathbf{V}}_1 &=& \mathbf{V}_1 \frac{1}{\varepsilon \mathbf{1} -
 \mathbf{E}_F} \mathbf{V}_2,\\
\tilde{\mathbf{V}}_2 &=& \mathbf{V}_3 \frac{1}{\varepsilon \mathbf{1} -
 \mathbf{E}_F} \mathbf{V}_2.
\end{eqnarray}
The next step in the recursion consists of the reduction of the blocks
with $\tilde{\mathbf{E}}_{F,2}$, and provides an homogeneous effective
Hamiltonian (panel $c$) with diagonal blocks and hopping matrices
renormalized by
\begin{eqnarray}
\tilde{\mathbf{E}}_F &=& \tilde{\mathbf{E}}_{F,1} + \tilde{\mathbf{V}}_1
 \frac{1}{\varepsilon \mathbf{1} - \tilde{\mathbf{E}}_{F,2}}
 \tilde{\mathbf{V}}_1^\dag + \tilde{\mathbf{V}}_2^\dag \frac{1}{\varepsilon
 \mathbf{1} - \tilde{\mathbf{E}}_{F,2}} \tilde{\mathbf{V}}_2,\\
\tilde{\mathbf{V}} &=& \tilde{\mathbf{V}}_1 \frac{1}{\varepsilon \mathbf{1} -
 \tilde{\mathbf{E}}_{F,2}} \tilde{\mathbf{V}}_2.
\end{eqnarray}
By repeating this process $M$ times, the self-energy correction to the
diagonal block $\mathbf{E}_F$ accounts for a system with $2^M$ layers
along the longitudinal direction. Therefore, the DOS reduces to
\begin{equation}
\mathcal{N}_0(\varepsilon) = -\frac{1}{\pi} \lim_{\eta \rightarrow 0}
 \sum_{i=1}^N \mathrm{Im} \bra{i,0} \frac{1}
 {(\varepsilon+i\eta)\mathbf{1} - \mathbf{E}_F-\mathbf{\Sigma}_F}
 \ket{i,0},
\label{eq:dos2}
\end{equation}
which only involves the inversion of a single block matrix.

\begin{figure}[tbp]
\includegraphics*[width=8.5cm]{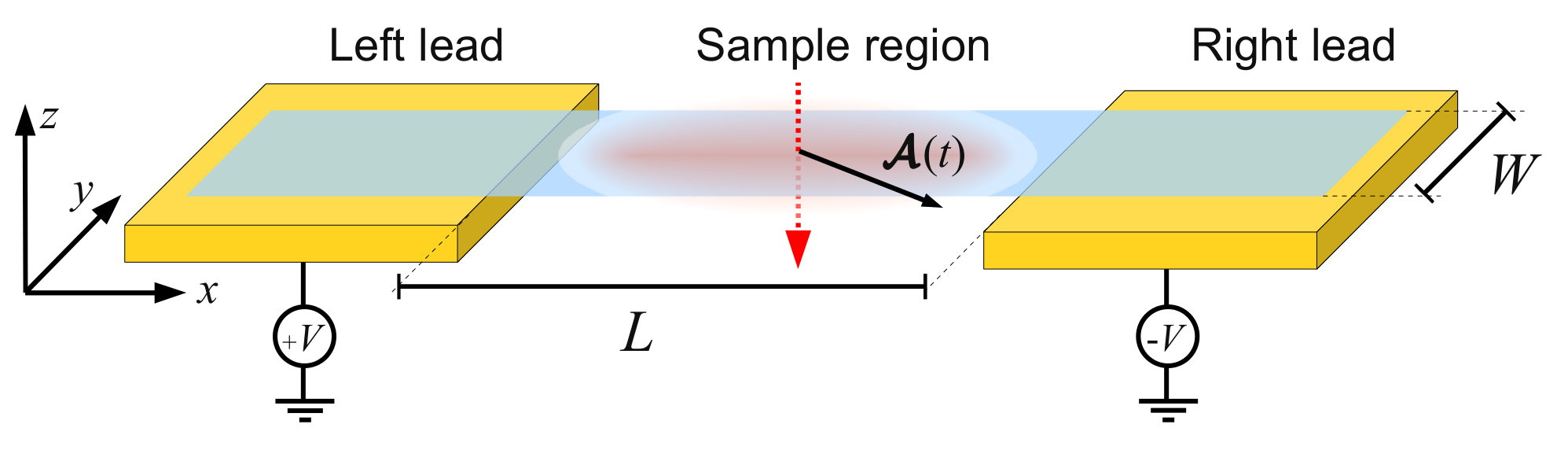}
\caption{(Color online) Scheme of the transport setup. The graphene
 nanoribbon is coupled to two metallic reservoirs. The central region
 (sample) is illuminated by a laser field perpendicular to the plane of
 the ribbon.}
\label{fig:device}
\end{figure}

The above scheme is also of great help for an efficient evaluation of
the dc component of the conductance. In this case we imagine a situation
as the one represented on figure \ref{fig:device}, where only a finite
region of space is affected by the laser (``the sample''). If we
consider the rest as reservoirs, then we can compute the Floquet-Green
functions for the sample by representing the $\alpha$-electrode ($\alpha
= L,R$) through a self-energy correction $\mathbf{\Sigma}_\alpha =
\mathbf{\Delta}_\alpha - i\mathbf{\Gamma}_\alpha$ as usual. How are
these Green's functions connected to the dc current? For non-interacting
electrons, the average (coherent) current over a period of the
modulation is given by: 
\begin{equation}
\bar{I}=\frac{2e}{h}\sum_{n}\int\left[  T_{R\leftarrow L}^{(n)}%
(\varepsilon)f_{L}(\varepsilon)-T_{L\leftarrow R}^{(n)}(\varepsilon
)f_{R}(\varepsilon)\right]  d\varepsilon,
\end{equation}
where $T_{R \leftarrow L}^{(n)}(\varepsilon)$ are the transmission
probabilities from the left ($L$) to the right ($R$) reservoirs
involving the emission/absorption of $n$ photons. These probabilities
can be written in terms of the Floquet-Green functions for the system
\cite{Kohler2005, Stefanucci2008}:
\begin{equation}
T_{R \leftarrow  L}^{(n)}(\varepsilon)= \Tr
\left[ 2 \mathbf{\Gamma}_{R,n}(\varepsilon)
\mathbf{G}_{(R,n)\leftarrow(L,0)}(\varepsilon) 
2 \mathbf{\Gamma}_{L,0}(\varepsilon)
\mathbf{G}^{\dag}_{(R,n)\leftarrow(L,0)}(\varepsilon)\right], 
\label{eq:transmittance}
\end{equation}
where $\mathbf{G}_{(R,n)\leftarrow(L,0)}(\varepsilon)$ is the block
matrix for the Floquet-Green function connecting the left and right
electrodes with the exchange of $n$ photons. Here the subindex $F$ was
omitted to simplify the notation. As a consequence of the
thermodynamical limit, the coupling to the (open) leads (see figure
\ref{fig:device}) implies a decay rate in the states within the sample
which is accounted by
\begin{equation}
\mathbf{\Gamma}_{\alpha,n}(\varepsilon)= - \mathrm{Im} \left[
\mathbf{\Sigma}_\alpha (\varepsilon+n\hbar\Omega) \right].
\end{equation}
Since we assume a laser affecting only the sample region, these terms can be
computed in terms of the ``bare'' self-energy correction
$\mathbf{\Sigma}_\alpha(\varepsilon)$ obtained in the absence of
time-dependent fields. The calculation of the current is thus completed
by assuming that the asymptotic occupation in the leads, where no
inelastic processes are allowed, is given by the usual Fermi functions
\begin{equation}
f_\alpha(\varepsilon) = \frac{1}{1+e^{\beta(\varepsilon-\mu_\alpha)}},
\end{equation}
where $\mu_\alpha$ is the electrochemical potential in the $\alpha$-lead
and $\beta$ the inverse temperature.

\section{Results}
\label{sec:results}

In this section we apply the discussed method to investigate
modifications in the electronic properties of graphene nanoribbons
induced by a laser field. In particular, we scrutinize the case
of a laser within the mid-infrared region of the spectrum, where photon
energies can be made smaller than the typical optical phonon energy 
while keeping the laser power small ($1 - 10 \unit{mW/\mu m^2}$). By an
appropriate tuning of the Fermi energy and polarization of the laser, we
show how one can dramatically change the electrical and transport
properties of the ribbons. This is illustrated for different sizes and
edge geometries of the ribbons.

\begin{figure}[tbp]
\includegraphics[width=8.5cm]{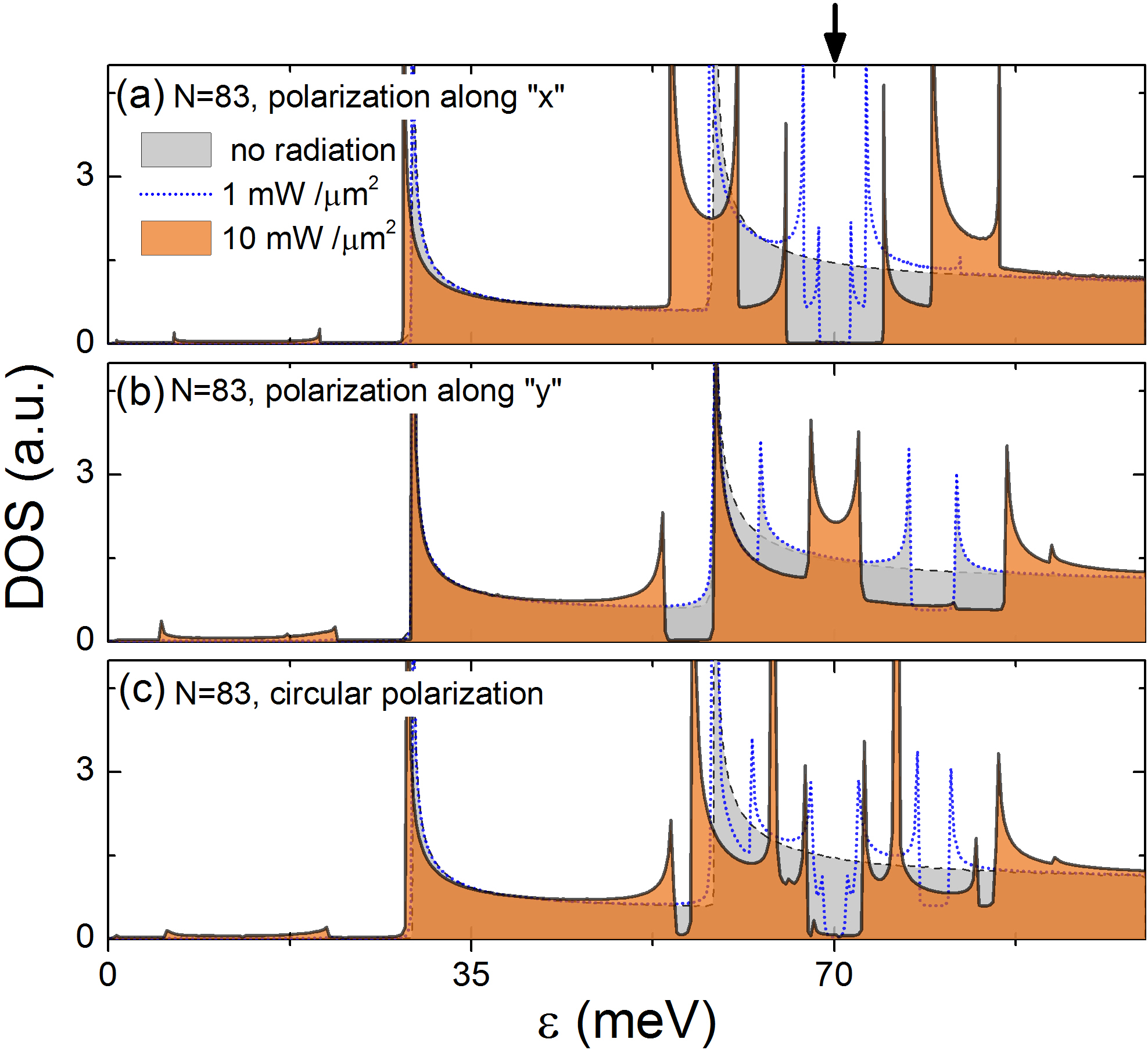}
\caption{(Color online) Average density of states for an armchair
 graphene nanoribbon with $N=83$ in the presence of linearly polarized
 radiation along the $x$ direction (a), $y$ (b) and circular
 polarization (c).}
\label{fig:dosn83}
\end{figure}

Figure \ref{fig:dosn83} shows the DOS (cf. equation \ref{eq:dos2}) for
an armchair ribbon with $N=83$ as a function of the Fermi energy. Three
cases are analyzed: (a) linear polarization along $x$, (b) linear
polarization along $y$, and (c) circular polarization. The energy of the
laser is fixed to $\hbar\Omega = 140\unit{meV}$, such that modifications in the
DOS are expected to occur in the vicinity of the energy $\varepsilon
\simeq 70\unit{meV}$ (marked by an arrow on the top panel). 

In panel $a$, we observe that the interaction with the laser leads to a
gap formation around $\varepsilon = \hbar \Omega/2$. The occurrence of
this gap is related to an inelastic backscattering process that enables
transitions between quasi-states in the Floquet spectrum. In this
picture, each mode contains a series of replicas accounting for
excitation states with different number of photons. In presence of
linearly polarized light along $x$, the electronic states are connected
to (photoexcited) hole states which belong to the same mode. Since
electron-hole symmetry is now preserved along the point $n \hbar \Omega$
for a replica with $n$ photons, the energy values at which the gap may
form are commensurate with $\hbar \Omega/2$. Additionally, the width of
the gap is shown to be sensitive to the transversal direction of the
momentum operator, such that it increases with the allowed values of
$k_y$. This can be observed in the structure of the DOS around
$\varepsilon = \hbar \Omega/2$, where two concentric gaps are developed,
each one of them related to a different mode.

In panel $b$, one can observe that the DOS is drastically changed. Here
the dynamical gap around $\varepsilon = \hbar \Omega/2$ is suppressed
and two ``satellite'' depletion regions, where the DOS is diminished,
appear instead. Depending on the particular normal mode affected by the
field, these depletions may or not develop as a true gap. For a laser power $P
= 10\unit{mW/\mu m^2}$ (solid line), the left depletion is crossed by a
van Hove singularity and a gap is opened in the region of the first
semiconductor band.

\begin{figure}[tbp]
\includegraphics[width=8.5cm]{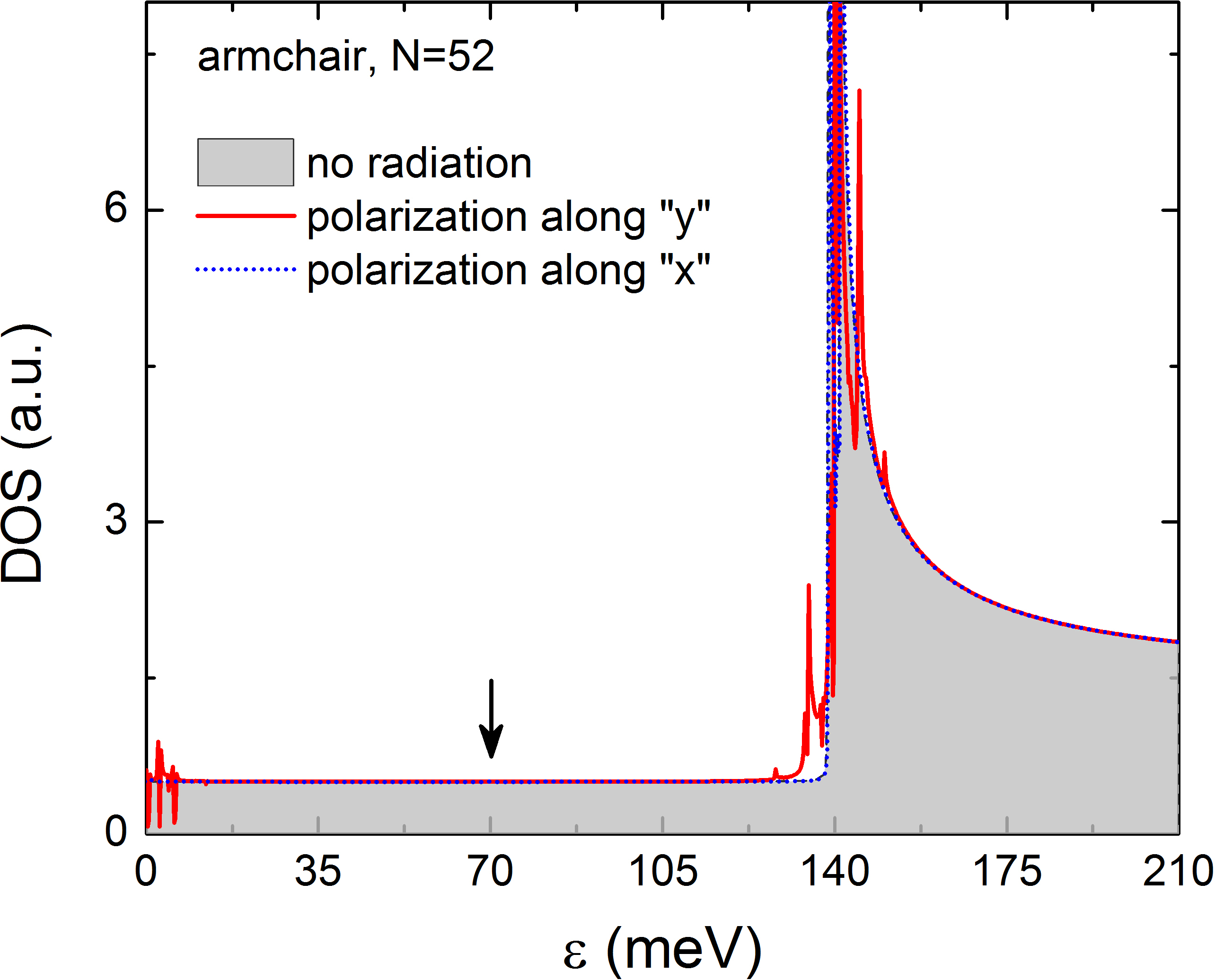}
\caption{(Color online) Average density of states for a metallic armchair
 nanoribbon $N=52$ in the presence of linearly polarized
 radiation ($\hbar \Omega = 140\unit{meV}, P = 1\unit{mW/\mu m^2}$)
 along the $y$ direction (solid red) and $x$ direction (dotted blue).}
\label{fig:metallic}
\end{figure}

Of particular interest is the case of a metallic armchair nanoribbon, as
shown in figure \ref{fig:metallic}. Here the DOS around $\varepsilon =
\hbar \Omega /2$ is completely immune to the influence of the laser,
such that no bandgaps appear regardless of the particular direction of
the polarization.


The above features observed in figures \ref{fig:dosn83}(a) and
\ref{fig:dosn83}(b) can be combined if the laser field points along any
intermediate direction between the $x$ and $y$ axes. This is shown in
panel $c$ where a circularly polarized field is considered. Although for
small samples this last case produces similar modifications in the DOS
as compared to linear polarization along the direction $\mathbf{x+y}$,
in the bulk limit these two cases become qualitatively different. 

\begin{figure}[tbp]
\includegraphics[width=8.5cm]{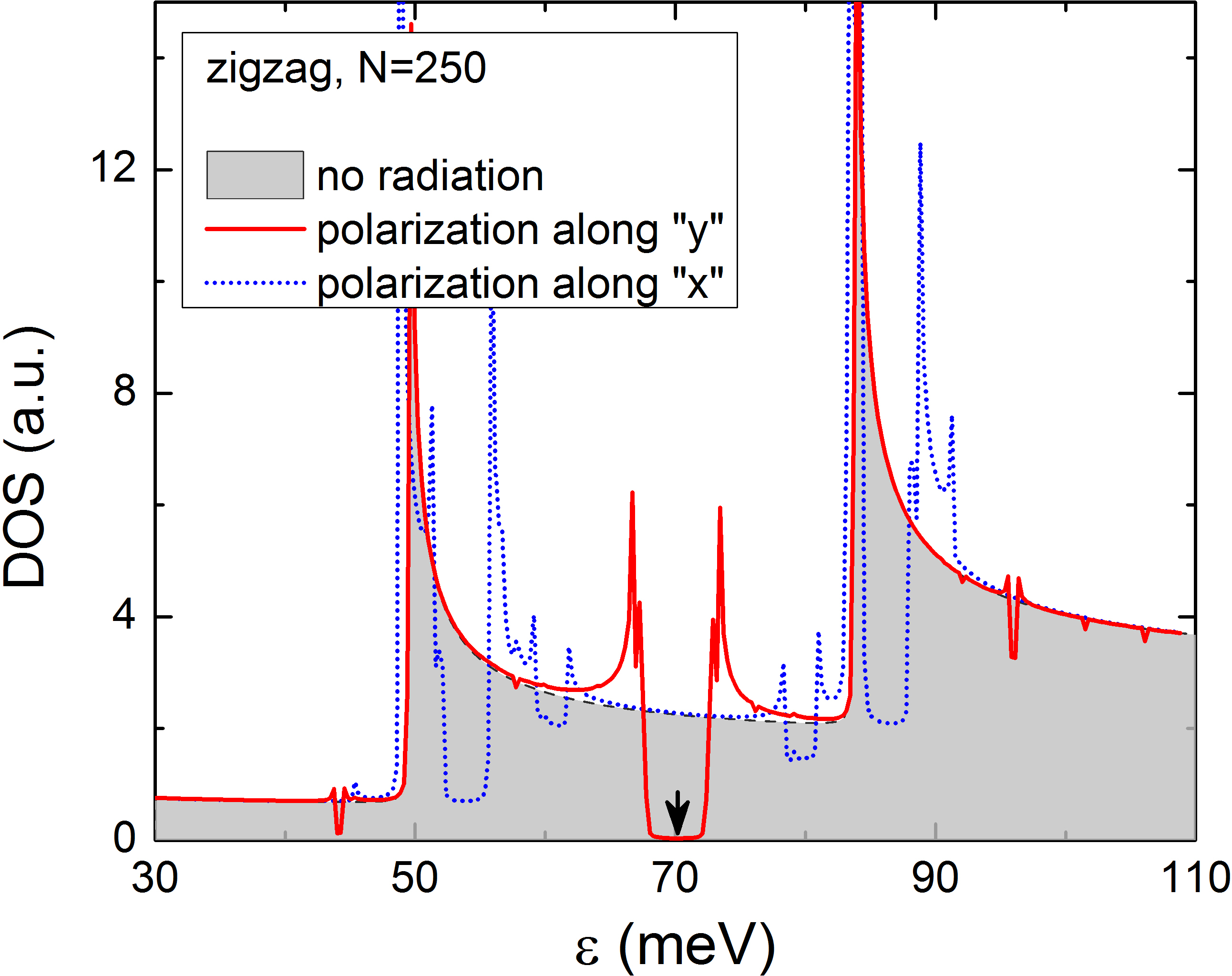}
\caption{(Color online) Average density of states for a zigzag
 nanoribbon with $N=250$ in the presence of linearly polarized radiation
 ($\hbar\Omega = 140\unit{meV}$) along the $y$-direction (solid red) and
 $x$-direction (dotted blue) for  a laser power $P = 1\unit{mW/\mu
 m^2}$.} 
\label{fig:doszz}
\end{figure}

As mentioned in section \ref{sec:results}, the developed technique is
not necessarily restricted to armchair ribbons and can be easily adapted
to other edge geometries. In figure \ref{fig:doszz} we show the DOS for
a zigzag nanoribbon with $N=250$. Compared to figure \ref{fig:dosn83},
the features in the DOS observed in the armchair case are also present
here. The difference now comes from the change in the relative angle
between the lattice orientation and the direction of the
polarization. In this sense, dynamical gaps around $\pm\hbar \Omega/2$
occur now for a laser along the $y$-direction. In addition, one can
observe small depletions at both sides of the gap which can be related
to the laser-induced coupling between different normal modes. We will
come back to this point below. By changing the polarization direction to
the $x$-axis, the dynamical gap is again suppressed and several
depletion regions appear.

An analysis of the bulk situation \cite{Oka2009,Calvo2011} would only
predict laser-induced depletions or gaps at $\pm n \hbar \Omega/2$ and
no dependence on the linear polarization direction. Natural questions
are therefore: why do these features away from $\pm n \hbar \Omega/2$
emerge? How do we reach the bulk limit? 

Two different kind of processes are at the heart of these phenomena:
intra-mode and inter-mode photon-induced transitions. For intra-mode
transitions, both the initial and the final
electronic states belong to the same transversal mode and the coupled states
are located symmetrically with respect to the charge neutrality point which leads
to the gaps or depletions at $\pm n \hbar \Omega/2$.

In contrast, inter-mode transitions couple states that are not
symmetrically located around the Dirac point. In armchair ribbons,
this is evident for the case of polarization in the $y$-direction, where
it turns out that inter-mode transitions are the only allowed
processes. Therefore, the polarization direction can be used to tune the
relative magnitude of the different kind of processes: intra and inter
mode.

\begin{figure}[tbp]
\includegraphics[width=8.5cm]{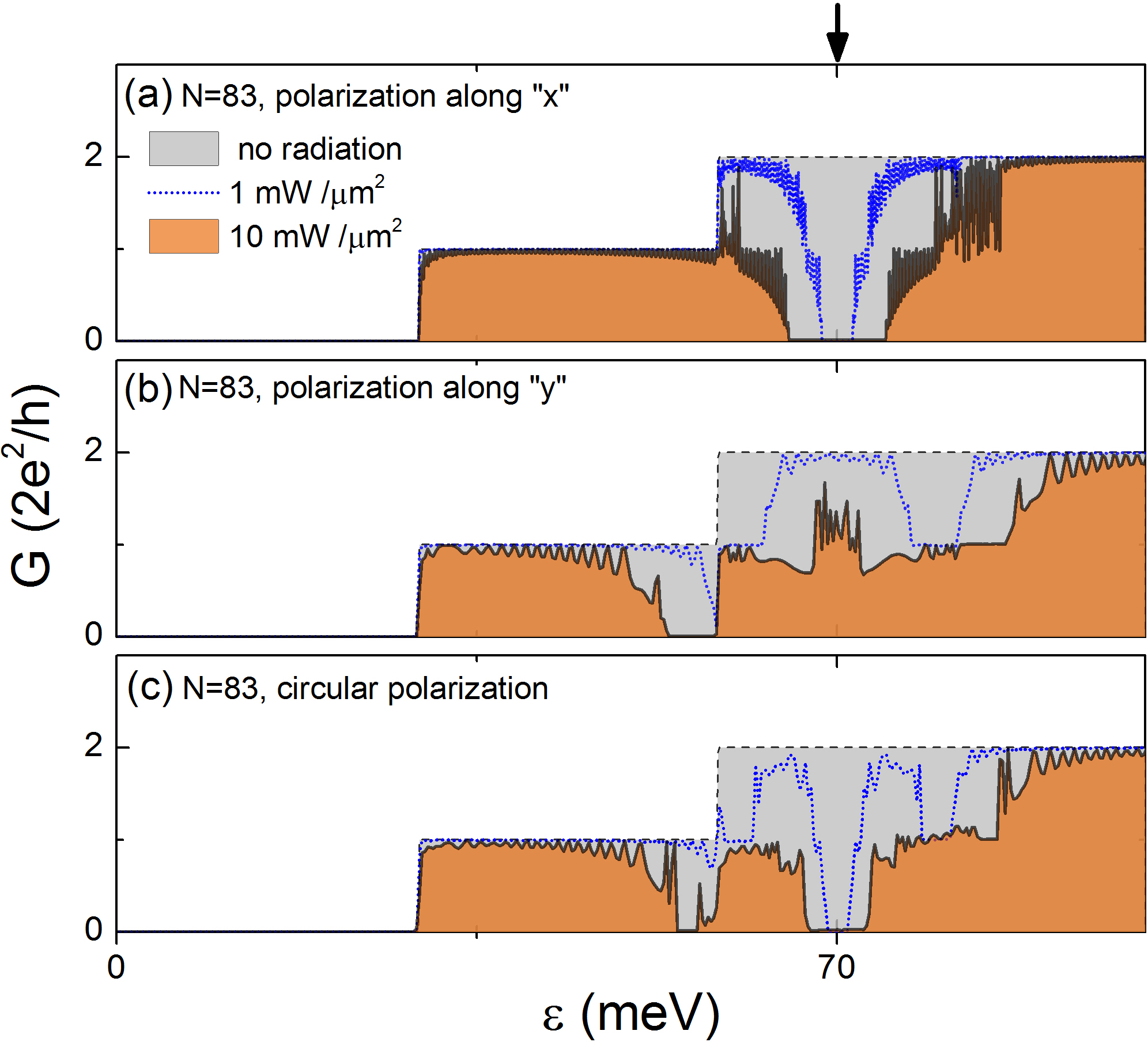}
\caption{(Color online) DC component of the conductance for an armchair
 graphene nanoribbon with $N=83$ in the presence of linearly polarized
 radiation along the $x$ direction (a), $y$ (b) and circular
 polarization (c).}
\label{fig:Gn83}
\end{figure}

We now calculate the dc component of the conductance at zero
 temperature. When $T_{R \leftarrow L}^{(n)}(\varepsilon)=T_{L
 \leftarrow R}^{(n)}(\varepsilon)$, it is straightforward to show
 (cf. equation \ref{eq:transmittance}) that the conductance reduces to:
\begin{equation}
G(\varepsilon) = \frac{2e^2}{h} \sum_n T_{R \leftarrow
 L}^{(n)}(\varepsilon).
\end{equation}
We use this expression, neglecting the small quantum pumping component
($\left|T_{R \leftarrow L}-T_{L \leftarrow R}\right| \ll T_{R \leftarrow L}$).  

Figure \ref{fig:Gn83} shows the conductance for the same set
of parameters as in figure \ref{fig:dosn83}. One observes that
the same qualitative features also appear in this case. By comparing the
panels in the figure one sees that the direction of the polarization may
be used as a ``knob'' to turn on or off the conductance at appropriate
values of the Fermi energy.

\begin{figure}[tbp]
\includegraphics[width=8.5cm]{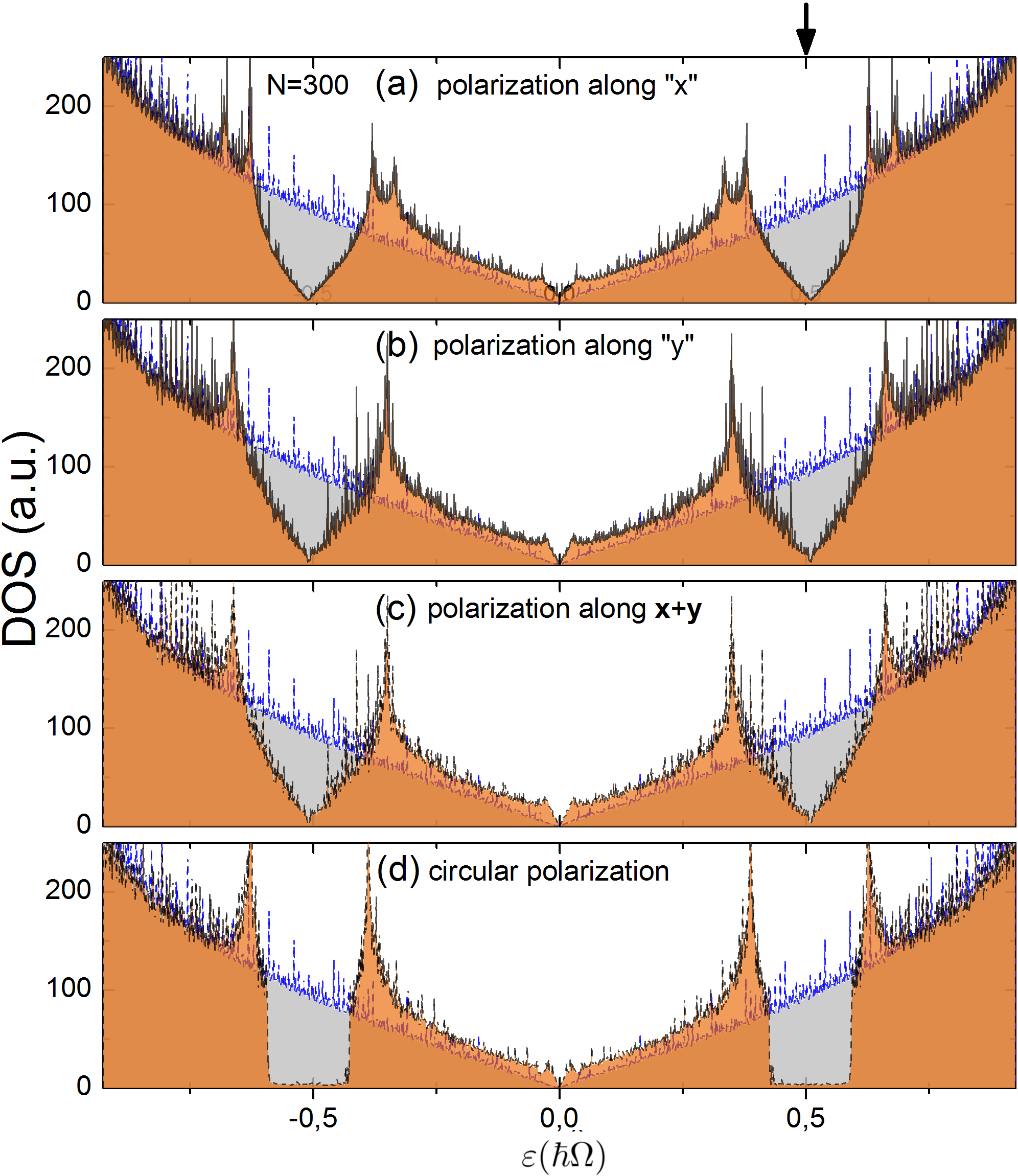}
\caption{(Color online) Average density of states for an armchair
 graphene nanoribbon with $N=300$ in the presence of linearly polarized
 radiation along the $x$ (a), $y$ (b) and $\mathbf{x+y}$ (c)
 directions. Circular polarization is shown in panel d. For a better
 visualization of the bulk limit a large value of $\Omega$ was taken,
 $\hbar \Omega = \gamma_0 = 2.7\unit{eV}$.}
\label{fig:dosn300}
\end{figure}

\subsection{Bulk limit}

Now that the origin of the features observed in figures \ref{fig:dosn83}
and \ref{fig:Gn83} has been clarified, we turn to the issue of how the
bulk limit is achieved. For a ribbon with a high number of modes,
a very large number of crossings in the Floquet spectrum become
``accumulated'' in the neighbourhood of $\pm n \hbar\Omega/2$. Once the
separation between these features becomes small enough, the depletions
merge and the system is insensitive to the particular direction of the
(linear) polarization. To illustrate the emergence of the bulk limit we
refer to figure \ref{fig:dosn300} showing the DOS for an armchair ribbon
with $N=300$. Linear polarization along the $x$, $y$ and $\bf{x}+\bf{y}$
directions, respectively, are considered in panels $a$,$b$ and $c$,
while circular polarization is shown in panel $d$. To account for a
large number of bands around the region $\hbar \Omega/2$, we increase
significantly the frequency and power of the laser to get a flavor of the
bulk effects while keeping the dimension of the Floquet space
treatable. A direct observation of panels $a$ to $c$ shows that the
sharp features observed in figures \ref{fig:dosn83} and \ref{fig:doszz}
for linearly polarized laser are now averaged in such a way that the
same DOS is obtained for any direction of the laser. On the other hand,
qualitative differences between linear and circular polarization become
evident in this limit, where one can see that a strong gap appears in
the last case.


\section{Conclusions}
\label{sec:conclusions}

We have presented a detailed account of a simulation
scheme for assessing the electrical properties (average conductance and
density of states) of laser-illuminated graphene nanoribbons. The scheme is based
on the application of Floquet theory to a realistic Hamiltonian which
allows the simulation of system sizes beyond the scope of the usual
{\bf \em{k.p}} approximation. The usefulness of this scheme is
illustrated on nanoribbons under different polarizations (linear and
circular). A simple analysis allowed us to rationalize the emergence of
the bulk 2d limit.

\end{document}